\documentclass[aps,prb,twocolumn,showpacs,superscriptaddress,groupedaddress,amsmath,amssymb]{revtex4-2}

\usepackage{graphicx}
\usepackage{dcolumn}
\usepackage{bm}
\usepackage{hyperref}
\hypersetup{
colorlinks = true, 
urlcolor = blue, 
linkcolor = blue, 
citecolor = blue
}

\usepackage{color}
\usepackage{ulem}

\begin{document}

\preprint{APS/123-QED}

\title{
Pair correlations of the hybridized orbitals in a ladder model \\ for the bilayer nickelate La$_3$Ni$_2$O$_7$
}
\author{
Masataka Kakoi,$^{1}$ Tatsuya Kaneko,$^{1}$ Hirofumi Sakakibara,$^2$ Masayuki Ochi,$^{1,3}$ and Kazuhiko Kuroki$^{1}$
}
\affiliation{
$^1$Department of Physics, Osaka University, Toyonaka, Osaka 560-0043, Japan\\
$^2$Advanced Mechanical and Electronic System Research Center(AMES), Faculty of Engineering, Tottori University, Tottori, Tottori 680-8552, Japan\\
$^3$Forefront Research Center, Osaka University, Toyonaka, Osaka 560-0043, Japan
}
\date{\today}
\begin{abstract}
To clarify the nature of high-temperature superconductivity in the bilayer nickelate La$_3$Ni$_2$O$_7$ under pressure, we investigate, using the density-matrix renormalization group method, the pair correlations in the two-orbital $t$-$J$ ladder model. 
While the interchain-intraorbital pair correlations exhibit a slow power-law decay in both orbitals, the {\it interorbital} pair correlation also develops strongly enough to be comparable with the intraorbital correlations.
These intra and interorbital pair correlations are enhanced by Hund's coupling, but more importantly, the interorbital pair correlation develops even when interorbital pairing glue mediated by Hund's coupling is absent.
Our finding suggests that the pair correlation in the present system develops as a hybridized two-orbital entity, which may have some implications on the superconductivity in the bilayer nickelate.
\end{abstract}

\maketitle

The recent discovery of high-temperature superconductivity under pressure with a $T_c$ of $\sim 80$~K in a bilayer Ruddlesden-Popper nickelate  La$_3$Ni$_2$O$_7$~\cite{Sun_2023} has initiated a new intensive wave of research in the field of condensed matter physics. 
Experimental reproductions that have followed the initial discovery have indeed established the occurrence of superconductivity in this material~\cite{Hou_replication_2023,Yanan-Zhang_replication_preprint,G-Wang_2024,Sakakibara_2024B,Wang_La2PrNi2O7_preprint,Y-Zhou_replication_preprint}. 
Also, already a large number of theoretical studies have appeared after the discovery of superconductivity~\cite{Luo_2023, QG-Yang_2023, Christiansson_2023, YF-Yang_2023, Y-Zhang_2023A, Y-Zhang_2023B, Lechermann_2023, W-Wu_2024, Y-Cao_2024, Chen_theory_preprint, YB-Liu_2023, C-Lu_2024,Y-Zhang_2024A, Oh_2023, Z-Liao_2023, K-Jiang_2024, Qin_2023,Tian_2024, DC-Lu_theory_preprint, R-Jiang_2024, Luo_theory_preprint, JX-Zhang_theory_preprint, Geisler_theory_preprint,C-Lu_theory_preprint2,Gu_theory_preprint,Y-Zhang_2024B,Kumar_theory_preprint,Ouyang_2024, H-Liu_theory_preprint, Ryee_theory_preprint,J-Chen_theory_preprint,Sakakibara_2024A,Kaneko_2024,Shen_2023,XZ-Qu_2024,Lange_2024,Schlomer_mixD_preprint,XZ-Qu_DMRG_preprint2}. 
Moreover, even the trilayer nickelate La$_4$Ni$_3$O$_{10}$ has been found to exhibit signatures of superconductivity under pressure with a lower $T_c$ of about $25$~K~\cite{Sakakibara_2024B,Q-Li_4310_2024, Y-Zhu_4310_preprint, M-Zhang_4310_preprint}, as expected theoretically~\cite{Sakakibara_2024B}.

Regarding the theories on La$_3$Ni$_2$O$_7$ that focus on the pairing mechanism, many of them agree on the point that the pairing involves interlayer nature, where the large interlayer hopping between the nearly half-filled $d_{3z^2-r^2}$ orbitals (or the interlayer magnetic exchange coupling induced by the interlayer hopping) plays an important role, which was a feature theoretically pointed out for this material in Ref.~\cite{Nakata_2017} by one of the present authors before the experimental discovery. 
In fact, nearly half-filled Hubbard (or $t$-$J$) model on a bilayer lattice~\cite{Nakata_2017, Kuroki_2002, Maier_2011, Mishra_2016, Maier_2019} or a two-leg ladder~\cite{Dagotto_1992, Troyer_1996,Dolfi_2015} has been known to be favorable for superconductivity for many years. 
However, in La$_3$Ni$_2$O$_7$, along with the nearly half-filled $d_{3z^2-r^2}$ orbitals, there exist nearly quarter-filled $d_{x^2-y^2}$ orbitals. 
The role played by the coupling between the $d_{3z^2-r^2}$ and $d_{x^2-y^2}$ orbitals, namely, the single-particle hybridization and the two-body interactions such as  Hund's coupling, and also, which one of the two orbitals dominates in the pairing, have been issues of debate. 

In Ref.~\cite{Sakakibara_2024A}, three of the present authors discussed the role played by those interorbital interactions using fluctuation exchange approximation, which is basically a weak coupling approach. 
On the other hand, we used the density-matrix renormalization group (DMRG) method~\cite{White_1992,White_1993,Schollwock_2011}, to study the interlayer pair correlations in a two-orbital two-leg Hubbard ladder that mimics the electronic structure of La$_3$Ni$_2$O$_7$ in a one-dimensional system (but without considering the interorbital two-body interactions)~\cite{Kaneko_2024}. 
There it was found that orbitals corresponding to the $d_{3z^2-r^2}$ and $d_{x^2-y^2}$ orbitals both exhibit slowly decaying correlations, even without Hund's coupling, with the former somewhat dominating in the decaying power. 
DMRG has also been adopted to investigate different types of models of La$_3$Ni$_2$O$_7$~\cite{Shen_2023,XZ-Qu_2024, XZ-Qu_DMRG_preprint2,Schlomer_mixD_preprint,Lange_2024}.
In terms of the two-orbital models, the numerical elucidation of the interplay of the $d_{3z^2-r^2}$ and $d_{x^2-y^2}$ orbitals is highly desired to approach the pairing mechanism in La$_3$Ni$_2$O$_7$. 

\begin{figure}[b]
  \includegraphics[width=0.95\linewidth]{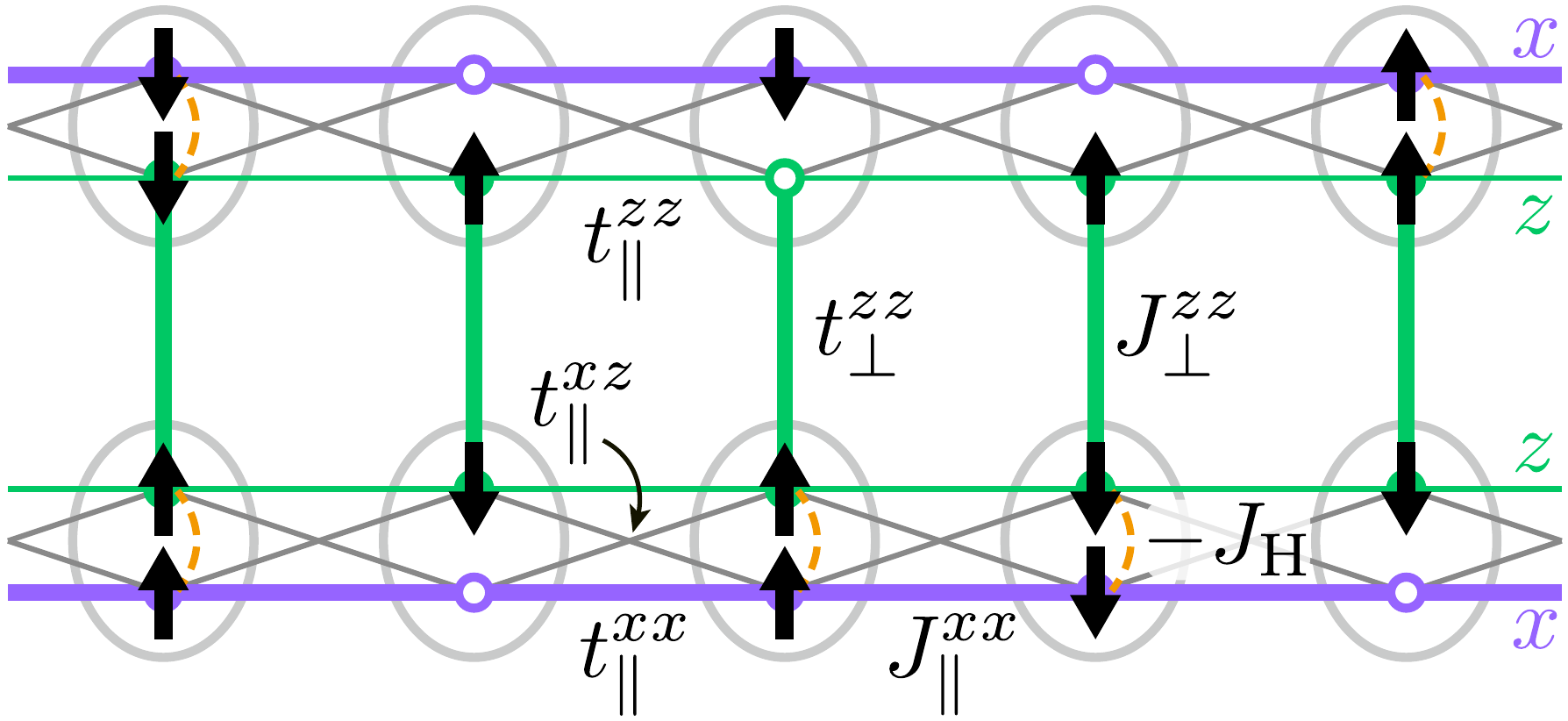}
  \caption{
  Two-orbital $t$-$J$ ladder at 3/8 filling.
  The $x$ and $z$ orbitals correspond to the $d_{3z^2-r^2}$ and $d_{x^2-y^2}$ orbitals, respectively, in the bilayer nickelate.
  }
  \label{fig:model}
\end{figure}

Given this background, to further investigate the effect of the interorbital interactions, here we study the pair correlations using DMRG in a two-orbital $t$-$J$ ladder that mimics La$_3$Ni$_2$O$_7$ in a similar manner as in Ref.~\cite{Kaneko_2024}, not only including the interlayer exchange coupling explicitly, but also considering Hund's coupling. 
We find that Hund's coupling encourages the correlations of the interchain pairs of both nearly half-filled (i.e., $d_{3z^2-r^2}$) and nearly quarter-filled (i.e., $d_{x^2-y^2}$) orbitals. 
More importantly, our calculation demonstrates that the correlation of the interorbital pairs exhibits a slow power-law decay, and this decaying behavior appears even without Hund's coupling.  
Our finding suggests that the hybridized orbital due to interorbital hopping (that exists in actual La$_3$Ni$_2$O$_7$) obtains the quasi-long-range superconducting correlation. 

To address the issues, we consider a two-orbital $t$-$J$ model [see Fig.~\ref{fig:model}], which is an effective model of the two-orbital Hubbard model in the strong coupling limit. 
Our $t$-$J$ model set in the ladder lattice prohibits the doubly occupied orbital at 3/8 filling, and the Hamiltonian $\hat{H} = \hat{H}_t + \hat{H}_J$ consists of the one-body term
\begin{align}\label{eq:Hamiltonian_TB}
    \hat{H}_t &= -\sum_{\mu,\nu}t_{\parallel}^{\mu\nu}\sum_{j,l}\sum_{\sigma}
    \left(
    \hat{\tilde{c}}^{\dag}_{j,l,\mu,\sigma}\hat{\tilde{c}}^{}_{j+1,l,\nu,\sigma} + {\rm H.c.}
    \right) 
    \nonumber\\
    &\quad -t_{\perp}^{zz}\sum_{j}\sum_{\sigma}
    \left(
    \hat{\tilde{c}}^{\dag}_{j,1,z,\sigma}\hat{\tilde{c}}^{}_{j,2,z,\sigma} + {\rm H.c.}
    \right) 
    \nonumber \\
    &\quad+\frac{\Delta E}{2}\sum_{j,l}\left(\hat{n}_{j,l,x}-\hat{n}_{j,l,z}\right)
\end{align}
and the spin interaction term
\begin{align}\label{eq:Hamiltonian_int}
    \nonumber
    \hat{H}_J &= J_{\parallel}^{xx}\sum_{j,l}\left(\hat{\bm{S}}_{j,l,x}\cdot\hat{\bm{S}}_{j+1,l,x} - \frac14 \hat{n}_{j,l,x}\hat{n}_{j+1,l,x}\right)\\
    \nonumber
    &\quad + J_{\perp}^{zz}\sum_{j}\left(\hat{\bm{S}}_{j,1,z}\cdot\hat{\bm{S}}_{j,2,z} - \frac14 \hat{n}_{j,1,z}\hat{n}_{j,2,z}\right)\\
    &\quad -2J_{\rm H}\sum_{j,l}\left(\hat{\bm{S}}_{j,l,x}\cdot\hat{\bm{S}}_{j,l,z} + \frac14 \hat{n}_{j,l,x}\hat{n}_{j,l,z}\right).
\end{align}
$\hat{\tilde{c}}^{}_{j,l,\mu,\sigma} = \hat{c}^{}_{j,l,\mu,\sigma}(1-\hat{n}_{j,l,\mu,\bar{\sigma}})$ is the projected annihilation operator of $\hat{c}_{j,l,\mu,\sigma}$ for an electron with spin $\sigma\ (=\uparrow,\downarrow)$ at site $j$ in chain $l$ $(=1,2)$, and orbital $\mu$ $(=x,z)$, where $\hat{n}_{j,l,\mu,\sigma} = \hat{c}^{\dag}_{j,l,\mu,\sigma} \hat{c}^{}_{j,l,\mu,\sigma}$ ($\hat{n}_{j,l,\mu}=\sum_{\sigma} \hat{n}_{j,l,\mu,\sigma}$) is the number operator and $\bar{\sigma}$ indicates the opposite spin of $\sigma$. 
Considering the bilayer nickelate system within a one-dimensional effective model, the orbitals $x$ and $z$ are associated with the $d_{x^2-y^2}$ and $d_{3z^3-r^2}$ orbitals, respectively. 
$\hat{\bm{S}}_{j,l,\mu} = (1/2)\sum_{\sigma,\sigma'} \hat{c}^{\dag}_{j,l,\mu,\sigma}\bm{\sigma}_{\sigma,\sigma'}\hat{c}^{}_{j,l,\mu,\sigma'}$ is the spin operator at site $j$ in chain $l$, and orbital $\mu$, where $\bm{\sigma}$ is a set of Pauli matrices $\bm{\sigma} = (\sigma^1, \sigma^2, \sigma^3)$.
$\Delta E$ ($>0$) is the energy difference between two orbitals, where the energy of the $x$ orbital is higher than the energy of the $z$ orbital, i.e., the $z$ ($x$) orbital becomes nearly half (quarter) filling. 
$t^{\mu\nu}_{\parallel}$ and $t^{\mu\nu}_{\perp}$ denote the intrachain and interchain hoppings, respectively.
$J_{\parallel}^{\mu\nu}$ and $J_{\perp}^{\mu \nu}$ indicate the intrachain and interchain spin-exchange couplings, respectively, and $J_{\rm H}$ ($>0$) is the Hund's (interorbital ferromagnetic) coupling. 

As for the interchain hopping $t^{\mu\nu}_{\perp}$, assuming that the overlap between two $d_{x^2-y^2}$ orbitals along the $z$ (rung) direction is small enough, we consider only interchain hopping $t^{zz}_{\perp}$ for the $d_{3z^2-r^2}$ orbital. 
In the high-symmetry structure (without tilt) of the bilayer nickelate under pressure, the interlayer hopping between the $d_{x^2-y^2}$ and $d_{3z^2-r^2}$ orbitals is zero, justifying $t_{\perp}^{xz}=0$. 
On the other hand, we take into account all intrachain hoppings. 
In this paper, we set $t^{xx}_{\parallel}=1$ as the energy unit and assume $t^{zz}_{\parallel}=0.25$ and $t^{xz}_{\parallel}=0.5$ to make a correspondence to the ratios of the intralayer hoppings estimated by the first-principle calculation in La$_3$Ni$_2$O$_7$~\cite{Sakakibara_2024A}. 
We use $t_{\perp}^{zz}=0.7$ and $\Delta E = 1$ employed in Ref.~\cite{Kaneko_2024} that suggests a good signature for superconductivity in the two-orbital Hubbard model.
The results with different values of $t_{\perp}^{zz}$ and $\Delta E$ are presented in the Supplemental Material~\cite{SM}. 

As for the spin interactions, we consider the intrachain antiferromagnetic coupling for the $x$ orbital $J_{\parallel}^{xx}$ ($>0$) and interchain antiferromagnetic coupling for the $z$ orbital $J_{\perp}^{zz}$ ($>0$). 
Since $J_{\parallel}^{zz}$ and $J_{\parallel}^{xz}$ are small relative to $J_{\parallel}^{xx}$ and $J_{\perp}^{zz}$ in the bilayer nickelate, we neglect $J_{\parallel}^{zz}$ and $J_{\parallel}^{xz}$ for simplicity. 
To comprehensively investigate the roles of the essential spin interactions in pairing properties within the two-orbital ladder model, we set $J_{\perp}^{zz}$ and $J_{\rm H}$ as variables while keeping $J_{\parallel}^{xx} = 0.5$. 
Assuming that the antiferromagnetic $J$ is on the order of $4t^2/U$ (where $U$ is the Hubbard repulsion), we have $J_{\parallel}^{xx}\simeq0.5$ and $J_{\perp}^{zz}\simeq0.25$ for $U=8$ with $t_{\parallel}^{xx}$ as the energy unit.
The ratio $J_{\perp}^{zz}/J_{\parallel}^{xx}$, however, could vary from $2$ due to unaccounted factors in the aforementioned estimation such as the interorbital repulsion $U'$ and the ligand $p$ orbital between the nickel ions. 

As shown in Fig.~\ref{fig:model}, the $z$ orbitals form the nearly half-filled ladder consisting of the strong interchain coupling ($t_{\perp}^{zz}$ and $J_{\perp}^{zz}$) and weak intrachain coupling ($t_{\parallel}^{zz}$). 
The electrons in the $x$ orbitals, which do not possess interchain couplings, are originally itinerant along the chain direction, while the interorbital hopping $t_{\parallel}^{xz}$ hybridizes the $x$ and $z$ networks. 
In addition, Hund's coupling $J_{\rm H}$ aligns the spins in the $x$ and $z$ orbitals within the single ion. 

\begin{figure}[tbp]
  \includegraphics[width=\linewidth]{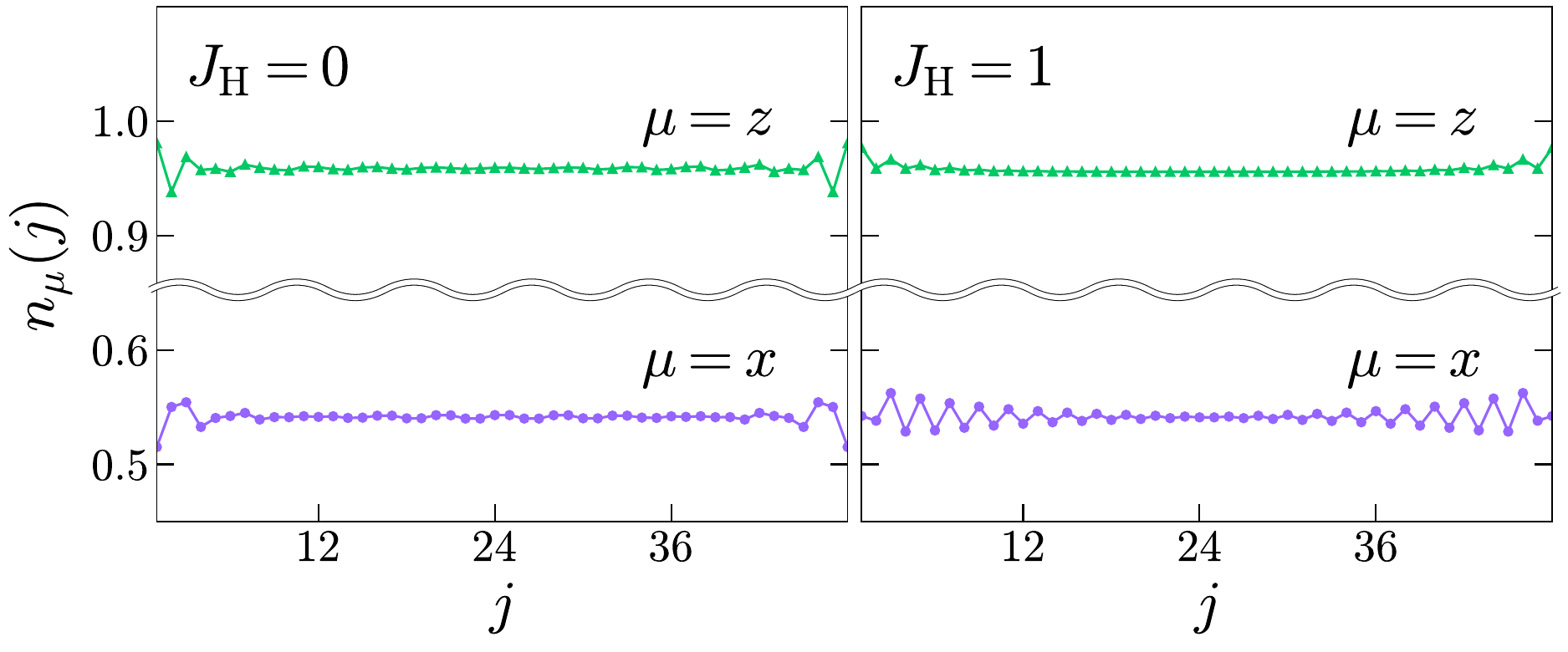}
  \caption{\label{fig:density}
  Local electron density $n_{\mu}(j) = 1/2\sum_l \langle\hat{n}_{j,l,\mu}\rangle$ at $J_{\rm H}=0$ (left panel) and $J_{\rm H}=1$ (right panel), where $J^{zz}_{\perp}=0.5$.
  }
\end{figure}

To compute the ground state of the two-orbital $t$-$J$ ladder, we employ the DMRG method implemented in the ITensor library~\cite{ITensor}. 
We carry out the DMRG calculations in ladders of lengths $L=48$ ($2\times48$ sites) with open boundary conditions. 
In this paper, we show the results at the bond dimension $m=10\,000$, where the truncation errors are on the order of at most $10^{-6}$.
We examine the $m$ and $L$ dependence of the results in the Supplemental Material~\cite{SM}. 
As shown in Fig.~\ref{fig:density}, the $z$~($x$) orbital is nearly half (quarter) filling in the ground state. 
Electron filling of each orbital is not significantly changed by Hund's coupling $J_{\rm H}$.
While $n_x(j)$ exhibits an oscillation when $J_{\rm H}=1$, the oscillations in the local electron density are small around the center of the ladder and a charge-density-wave character is not substantial.
To explore the nature of superconductivity in the two-orbital $t$-$J$ ladder, we calculate the pair correlation function $P_{\perp}^{\mu\mu}(r)=\langle\hat{\Delta}_{j,\mu\mu}^{\dag}\hat{\Delta}^{}_{j+r,\mu\mu}\rangle$, where $\hat{\Delta}_{j,\mu\mu} = (\hat{\tilde{c}}_{j,1,\mu,\uparrow}\hat{\tilde{c}}_{j,2,\mu,\downarrow}-\hat{\tilde{c}}_{j,1,\mu,\downarrow}\hat{\tilde{c}}_{j,2,\mu,\uparrow})/\sqrt{2}$ is the interchain spin-singlet pair annihilation operator on orbital $\mu$  ($=x,z$) at site $j$.
Here, we show the pair correlation function for the reference site $j=j_{\rm ref}=L/4$. 

\begin{figure}[tbp]
  \includegraphics[width=\linewidth]{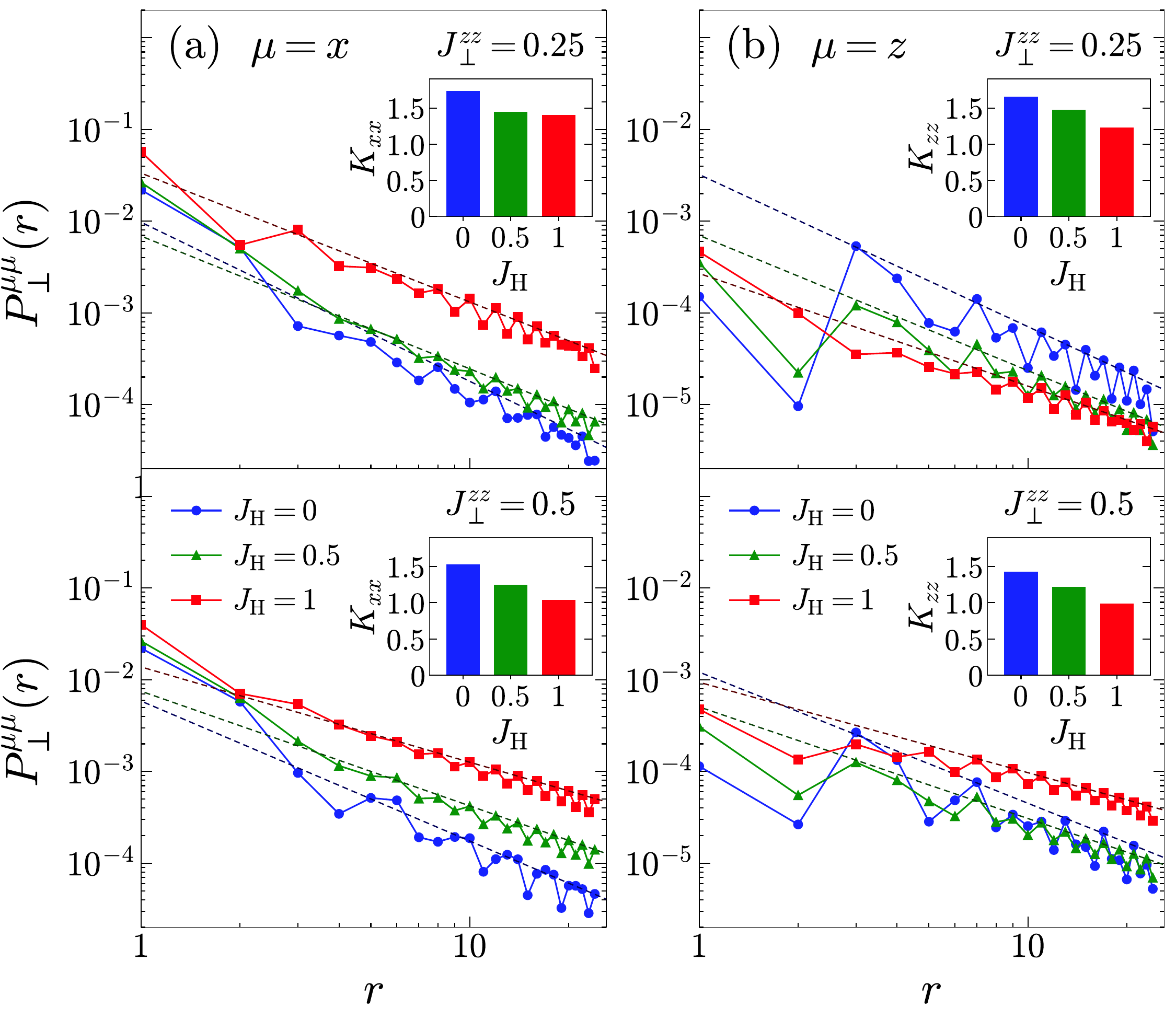}
  \caption{
  Pair correlation functions $P_{\perp}^{\mu\mu}(r)$ for various values of Hund's coupling $J_{\rm H}$.
  (a)~$P_{\perp}^{xx}(r)$ at $J_{\perp}^{zz} = 0.25$ (upper panel) and $J_{\perp}^{zz} = 0.5$ (lower panel).  
  (b)~$P_{\perp}^{zz}(r)$ at $J_{\perp}^{zz} = 0.25$ (upper panel) and $J_{\perp}^{zz} = 0.5$ (lower panel). 
  The insets show the decay exponents of $P_{\perp}^{\mu\mu}(r)$, where the exponent $K_{\mu\mu}$ is extracted by fitting the crests of the data points at $r \ge 6$.
  }
  \label{fig:pair-corr_Hund-dependence}
\end{figure}

In Fig.~\ref{fig:pair-corr_Hund-dependence}, we compare the pair correlation functions $P_{\perp}^{\mu\mu}(r)$ for various values of $J_{\perp}^{zz}$ and $J_{\rm H}$. 
We find that the pair correlations of both orbitals exhibit power-law decays ($P_{\perp}^{\mu\mu}(r)\propto r^{-K_{\mu\mu}}$), as is consistent with the behavior in the two-orbital Hubbard ladder~\cite{Kaneko_2024}.
Reflecting the presence of many carriers in the $x$ orbitals close to quarter filling, $P_{\perp}^{xx}(r)$ is larger than $P_{\perp}^{zz}(r)$ in the range we plotted. 
In one-dimensional systems, on the other hand, the correlations persisting over long distances are also crucial, and therefore we show the decay exponent $K_{\mu\mu}$ in the inset of Fig.~\ref{fig:pair-corr_Hund-dependence}. 
Here, a smaller $K_{\mu\mu}$ is preferable to a quasi-long-range order (i.e., slower decay of the pair correlation). 
As seen in Fig.~\ref{fig:pair-corr_Hund-dependence}(b), the decay of $P_{\perp}^{zz}(r)$ at $J_{\perp}^{zz}=0.5$ is slower (i.e., has smaller $K_{zz}$) than that at $J_{\perp}^{zz}=0.25$. 
This tendency is consistent with the case in the one-orbital $t$-$J$ ladder, in which a larger $J_{\perp}$ is favorable for the pair formation~\cite{Dagotto_1992,Troyer_1996}. 
Moreover, our calculations in the two-orbital $t$-$J$ ladder show that Hund's coupling $J_{\rm H}$ enhances the pair correlations at long distances, supporting a smaller decay exponent $K_{zz}$. 
$P_{\perp}^{xx}(r)$ in Fig.~\ref{fig:pair-corr_Hund-dependence}(a) also shows a similar decay tendency against $J_{\perp}^{zz}$ and $J_{\rm H}$. 
As summarized in the insets of Fig.~\ref{fig:pair-corr_Hund-dependence}, we find that larger $J_{\rm H}$ as well as larger $J_{\perp}^{zz}$ makes $K_{\mu\mu}$ smaller for both orbitals, i.e., they promote the quasi-long-range superconducting order.
Arbitrariness in the choice of data points used for fitting and the choice of the reference site may affect the results for $K_{\mu\mu}$. 
We confirm that the $J_{\perp}^{zz}$ and $J_{\rm H}$ dependence of $K_{\mu\mu}$ gives a similar tendency to Fig.~\ref{fig:pair-corr_Hund-dependence} even when different fitting procedures or averaged pair correlations are used (see the Supplemental Material~\cite{SM}).
While $K_{xx} > K_{zz}$ in most of the parameter sets used in Fig.~\ref{fig:pair-corr_Hund-dependence}, $K_{xx}$ is comparable to $K_{zz}$, suggesting that both orbitals cooperatively contribute to the pairing. 

\begin{figure}[t]
  \includegraphics[width=\linewidth]{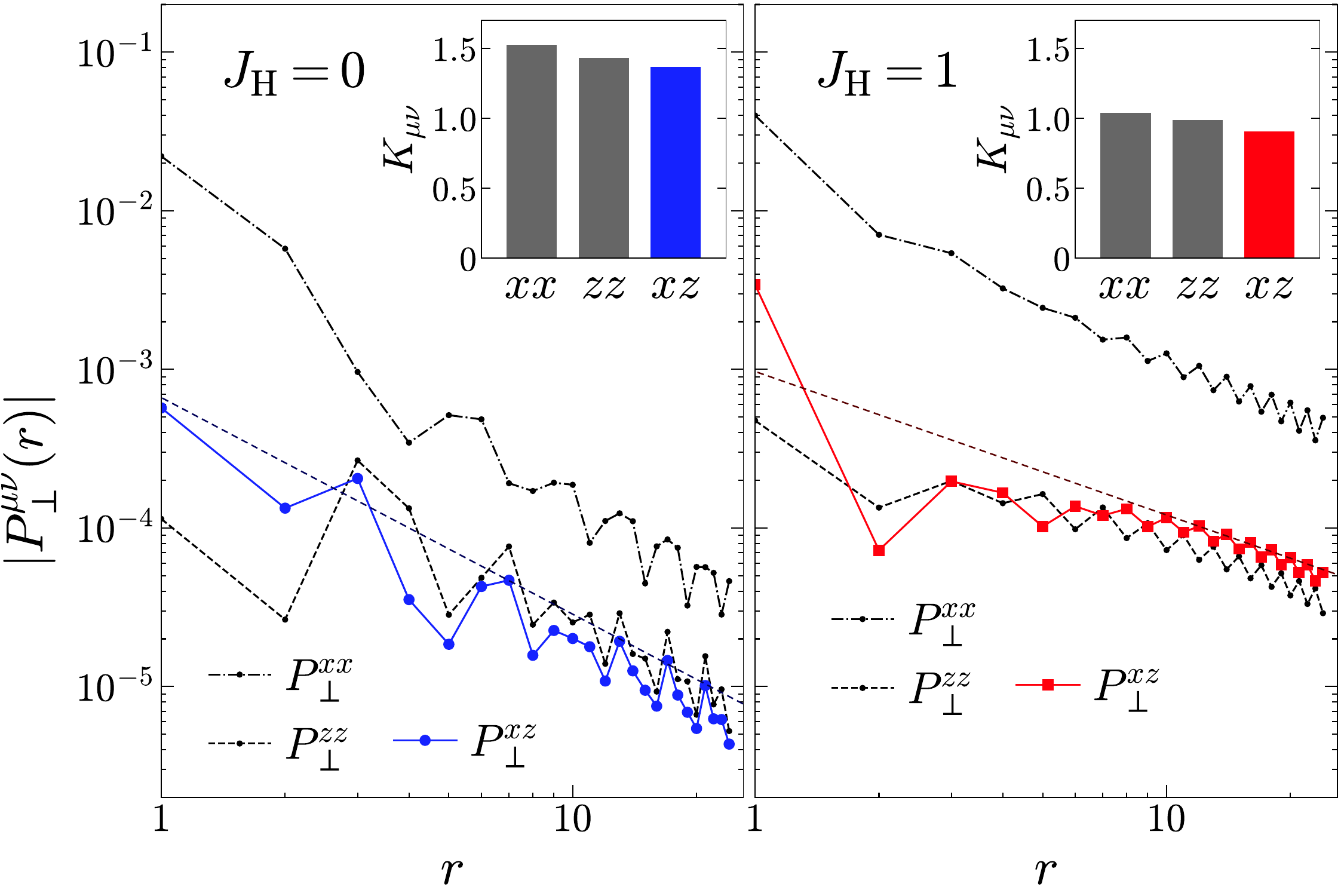}
  \caption{
  Interorbital pair correlation functions $P_{\perp}^{xz}(r)$ at $J_{\rm H}=0$ (left panel) and $J_{\rm H} = 1$ (right panel), where $J_{\perp}^{zz}=0.5$. 
  Intraorbital pair correlation functions $P_{\perp}^{\mu\mu}(r)$ are also presented for comparison.
  The insets show the decay exponents of $P_{\perp}^{xz}(r)$ denoted by $K_{xz}$ with $K_{xx}$ and $K_{zz}$.
  The exponent $K_{xz}$ is extracted by fitting the crests of the data points at $r\geq6$.
  }
 \label{fig:pair-corr_xz}  
\end{figure}

Curiously, $P_{\perp}^{xx}(r)$ exhibits a comparable power-law decay with $P_{\perp}^{zz}(r)$ even at $J_{\rm H} = 0$ (and $J_{\perp}^{xx}=0$). 
This indicates that the orbital hybridization via $t^{xz}_{\parallel}$ is a crucial factor for the pair correlation of the $x$ component at $J_{\rm H}=0$ because the interorbital coupling $J_{\rm H}$ $(=0)$ does not create the local interchain spin correlation between the $x$ orbitals, as we shall see explicitly later.  
A developed $x$-component pair correlation without $J_{\rm H}$ is also seen in the two-orbital Hubbard ladder~\cite{Kaneko_2024}. 
The present result is even more curious than in the case of the Hubbard ladder because the intrachain-interorbital exchange coupling $J^{xz}_{\parallel}$, 
which is proportional to $\sim (t^{xz}_{\parallel})^2/U$ in the Hubbard ladder (at $U \gg \Delta E , t^{xz}_{\parallel}$) and can induce the interchain $x$-$x$ spin correlation through $z$-$z$ spin correlation, is absent in the present model. 
Here, to examine the interorbital contribution to the pairing more directly, we compute the correlation of the interchain-interorbital spin singlet pair described by 
$\hat{\Delta}_{j,xz}^{} = (\hat{\tilde{c}}^{}_{j,1,x,\uparrow}\hat{\tilde{c}}^{}_{j,2,z,\downarrow}-\hat{\tilde{c}}^{}_{j,1,x,\downarrow}\hat{\tilde{c}}^{}_{j,2,z,\uparrow})/\sqrt{2}$ and present its correlation function $P_{\perp}^{xz}(r)$ in Fig.~\ref{fig:pair-corr_xz}. 
The decay of $P_{\perp}^{xz}(r)$ is comparable to that of $P_{\perp}^{xx}(r)$ and $P_{\perp}^{zz}(r)$ at $J_{\rm H} = 0$, suggesting that the $x$-$z$ singlet pair also strongly contributes to the superconducting correlation. 
The decay exponent $K_{xz}$ is presented in the inset of Fig.~\ref{fig:pair-corr_xz}, where $K_{xz}$ is the smallest and comparable to $K_{xx}$ and $K_{zz}$.
Even if we extract $K_{\mu\nu}$ from the averaged pair correlation $\bar{P}_{\perp}^{\mu\nu}(r)$, we find a small decay exponent for the $x$-$z$ pair (see the Supplemental Material~\cite{SM}). 
Our numerical demonstration implies that the interorbital component is also a considerable ingredient for the pairing in the presence of $t_{\parallel}^{xz}$. 
A slow decay of $P_{\perp}^{xz}(r)$ also appears at $J_{\rm H} = 1$ and the decay exponent $K_{xz}$ is still the smallest, suggesting the significance of the $x$-$z$ component of the pair regardless of $J_{\rm H}$. 

\begin{figure}[t]
  \includegraphics[width=\linewidth]{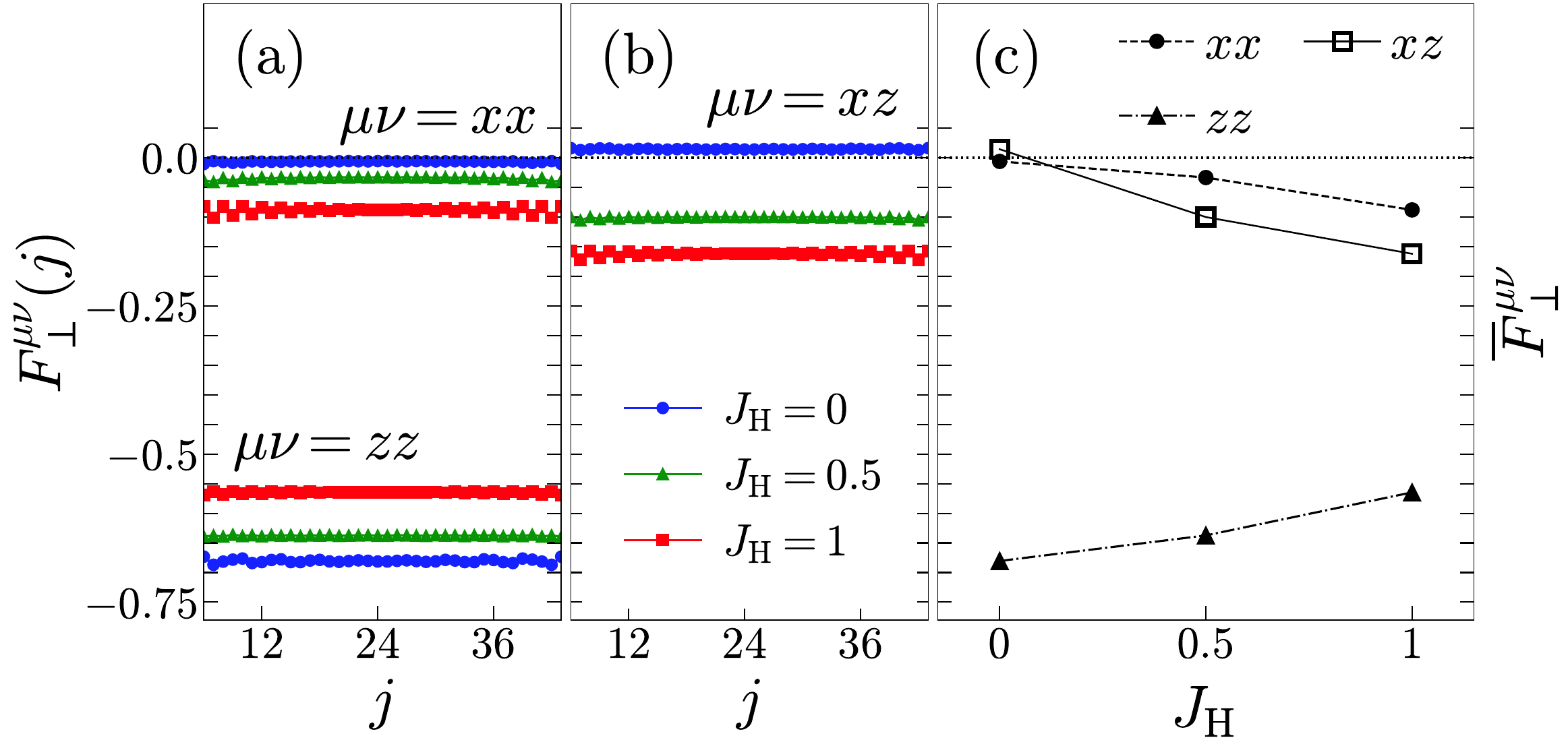}
  \caption{Interchain spin correlation functions $F_{\perp}^{\mu\nu}(j) = \langle \hat{\bm{S}}_{j,1,\mu}\cdot\hat{\bm{S}}_{j,2,\nu} \rangle$ for various values of Hund's coupling $J_{\rm H}$ (where $J_{\perp}^{zz}=0.5$). 
  (a) Intraorbital components $F_{\perp}^{xx}(j)$ and $F_{\perp}^{zz}(j)$.
  (b) Interorbital component $F_{\perp}^{xz}(j)$.
  (c) $J_{\rm H}$ dependence of the averaged spin correlation $\bar{F}_{\perp}^{\mu\nu}$, where $F_{\perp}^{\mu\nu}(j)$ is averaged over the sites from $j=12$ to $j=36$.  
  }
 \label{fig:spin-corr}  
\end{figure}

To understand the underlying spin structure, we present the local interchain spin correlation $F_{\perp}^{\mu\nu}(j) = \langle \hat{\bm{S}}_{j,1,\mu}\cdot\hat{\bm{S}}_{j,2,\nu} \rangle$ and its average $\bar{F}_{\perp}^{\mu\nu}$ in Fig.~\ref{fig:spin-corr}. 
At $J_{\rm H}=0$, $F_{\perp}^{zz}(j)$ is close to the value of the ideal spin-singlet ($=-0.75$) because of $J^{zz}_{\perp}$ that directly forms the spin-singlet, whereas $F_{\perp}^{xx}(j)$ and $F_{\perp}^{xz}(j)$ are nearly zero reflecting $J^{xx}_{\perp} = J^{xz}_{\perp}=0$. 
At large $J_{\rm H}$, on the other hand, antiferromagnetic correlations in $F_{\perp}^{xx}(j)$ and $F_{\perp}^{xz}(j)$ are enhanced by $J_{\rm H}$, implying that the effective $x$-$x$ and $x$-$z$ spin couplings are generated by the combination of $J^{zz}_{\perp}$ and $J_{\rm H}$ as pointed out by the previous studies~\cite{C-Lu_2024,Oh_2023}.  
While the $z$-$z$ component is suppressed by $J_{\rm H}$, its magnitude is still the largest. 
Hence, the glue of the $z$-$z$ pair is active even at larger $J_{\rm H}$. 

The enhancement of the interchain $x$-$x$ and $x$-$z$ spin-singlet correlations ($F_{\perp}^{xx}$ and $F_{\perp}^{xz}$) upon increasing $J_{\rm H}$ is consistent with the enhancement of the $x$-$x$ and $x$-$z$ pair correlations ($P_{\perp}^{xx}$ and $P_{\perp}^{xz}$) seen in Figs.~\ref{fig:pair-corr_Hund-dependence} and \ref{fig:pair-corr_xz} as $J_{\rm H}$ is increased. 
On the other hand, there are some contrasting features between the interchain spin correlations and pair correlations. 
First, at $J_{\rm H}=0$, both $x$-$z$ and $x$-$x$ spin correlations are very small, which is naturally expected in the absence of $J_{\perp}^{xx}$, $J_{\perp}^{xz}$, and $J^{xz}_{\parallel}$. 
This is in striking contrast to the fact that even at $J_{\rm H}=0$, the interchain $x$-$z$ and $x$-$x$ pair correlations exhibit a slow decay. 
In other words, quasi-long-range interchain pair correlation develops in the $x$-$z$ and $x$-$x$ channel {\it even in the absence of pairing glues mediated by $J_{\perp}^{zz}$ and $J_{\rm H}$}~\cite{C-Lu_2024,Oh_2023}. 
This suggests that in the presence of $t^{xz}_{\parallel}$, the pairs must be described by $x$-$z$ {\it hybridized entities}, where the pairing glue fundamentally originates from the strong interchain exchange coupling $J^{zz}_\perp$ of the nearly half-filled $z$ orbitals, but $x$-$z$ and $x$-$x$ interchain pair correlations are also comparably strong. 
Second, although $J_{\rm H}$ reduces the spin-singlet correlation of the $z$-$z$ component [see Fig.~\ref{fig:spin-corr}], $J_{\rm H}$ enhances the pair correlation $P_{\perp}^{zz}(r)$ [see Fig.~\ref{fig:pair-corr_Hund-dependence}].  
This may also support our picture that the pairs should be described by $x$-$z$ hybridized entities in the presence of $t^{xz}_{\parallel}$. 
Namely, the enhancement of the $x$-$z$ pair correlation with increased $J_{\rm H}$ results in an enhanced pair correlation of the $x$-$z$ hybridized entity as a whole, and hence leads to the enhancement of the $z$-$z$ pair correlation. 

Since the hybridization due to $t^{xz}_{\parallel}$ gives the nonlocal effects, an interpretation of the pair in real space is nontrivial.  
The optimal definition of the local pair and examination of its pair correlation in strongly correlated and hybridized two-orbital systems is an important open issue. 
We must also note that the hybridization effect in one-dimensional systems is strong relative to the actual two-dimensional La$_3$Ni$_2$O$_7$, in which the hybridization between the $d_{x^2-y^2}$ and $d_{3z^2-r^2}$ orbitals vanishes along the $k_x=\pm k_y$ line on the square lattice~\cite{Sun_2023}. 
Hence, our idea for the ladder system may potentially overestimate the effect of $t_{\parallel}^{xz}$ in the actual two-dimensional bilayer nickelate. 
Also, we considered only Hund's coupling $J_{\rm H}$ as the interorbital two-body interaction. 
The effect of other interorbital interactions such as the interorbital repulsion $U'$ or the pair hopping $J_{\rm pair}$ remains an open issue. 
In fact, if we apply the fluctuation exchange approximation, which is basically a weak coupling approach, to a three-dimensional model of La$_3$Ni$_2$O$_7$, we find that while $J_{\rm H}$ alone does enhance superconductivity within a realistic parameter range, both $U'$ and $J_{\rm pair}$ suppress superconductivity~\cite{Sakakibara_unpublished} so that the two-body interorbital interactions in total result in a slight suppression of superconductivity~\cite{Sakakibara_2024A}. 

To summarize, we have investigated the pair correlations using DMRG in a two-orbital $t$-$J$ ladder including  Hund's coupling that mimics La$_3$Ni$_2$O$_7$. 
Our calculation demonstrates that the correlation of the interorbital $x$-$z$ pairs exhibits a slow power-law decay as well as the $x$-$x$ and $z$-$z$ pairs, and they are promoted by Hund's coupling $J_{\rm H}$.  
Our numerics suggest that the hybridized entity due to the interorbital hopping $t_{\parallel}^{xz}$ obtains the quasi-long-range superconducting correlation.  
The necessity of such a picture for describing the pairing state in the two-orbital ladder system may have some implications on the superconductivity in the bilayer nickelate.

\begin{acknowledgments}
This work was supported by Grants-in-Aid for Scientific Research from JSPS, KAKENHI Grants No. JP20H01849, No. JP24K06939 (T.K.), No. JP22K03512 (H.S.), No. JP22K04907 (K.K.), and No. JP24K01333. M.K. was supported by Program for Leading Graduate Schools: Interactive Materials Science Cadet Program and by Kato Foundation for Promotion of Science, Grant No. KS-3614. The computing resource is supported by the supercomputer system (system-B) in the Institute for Solid State Physics, the University of Tokyo, and the supercomputer of Academic Center for Computing and Media Studies (ACCMS), Kyoto University. The DMRG calculations were performed using the ITensor library~\cite{ITensor}. 
\end{acknowledgments}

{\it Note added.---}
Recently, we became aware of another theoretical study that performs DMRG calculations in a $t$-$J$ model~\cite{XZ-Qu_DMRG_preprint2} during the finalization process of the present study. 
The model studied there is similar to ours, and the tendency of the pair correlation against Hund's coupling is consistent while the different parameter regimes are studied. 
Besides, we studied the interorbital pair correlations, which were not studied in this, or any other previous studies.

\bibliography{refs}

\clearpage
\onecolumngrid
\setcounter{figure}{0}
\setcounter{equation}{0}
\renewcommand{\thefigure}{S\arabic{figure}}

\section*{Supplemental Material}

\subsection*{A. Bond dimension $m$ and length $L$ dependence of pair correlation functions}

\begin{figure}[htbp]
  \includegraphics[width=0.65\linewidth]{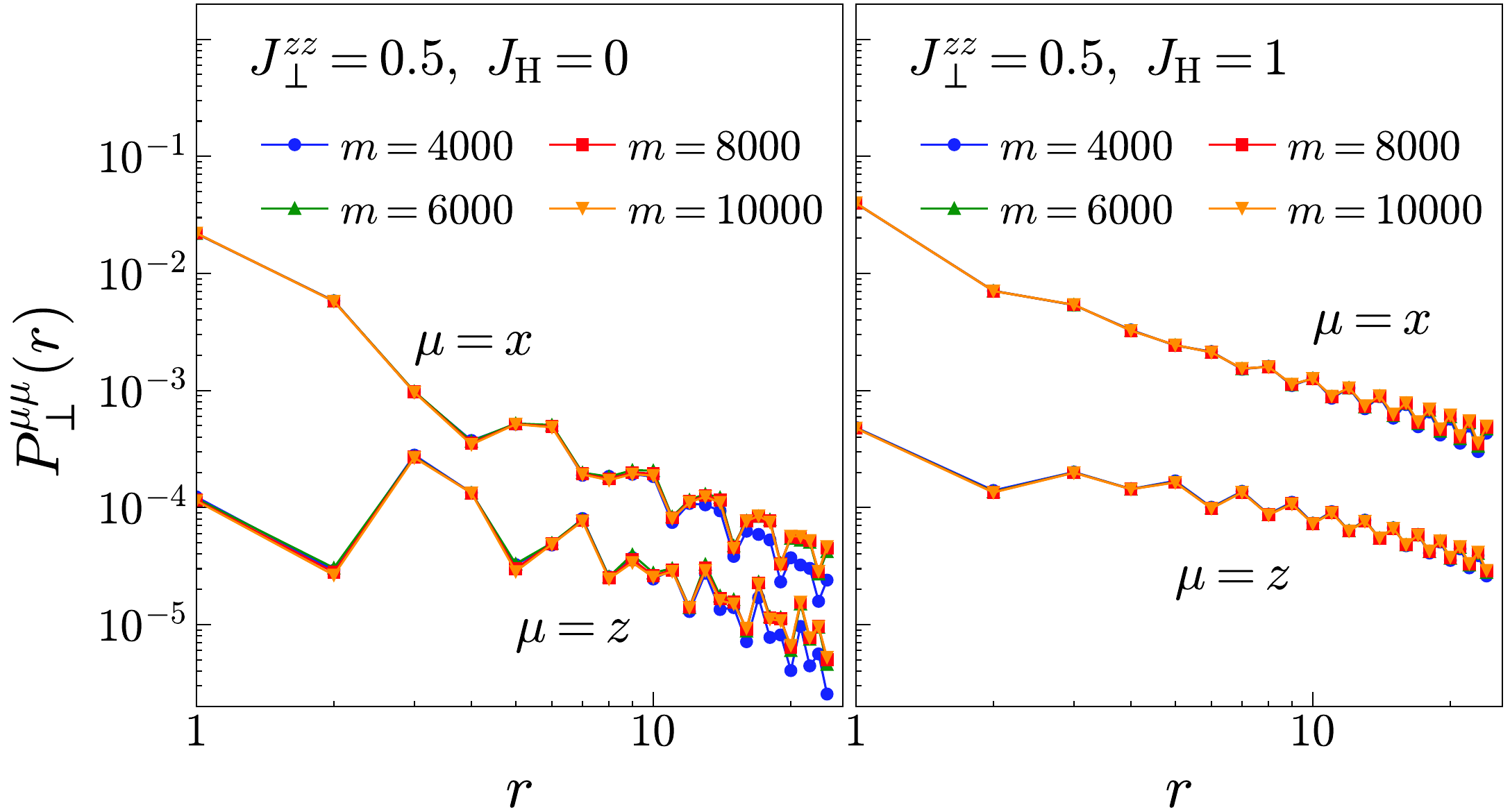}
  \caption{\label{fig:pair-corr_m-dependence}
  $m$ dependence of $P_{\perp}^{xx}(r)$ and $P_{\perp}^{zz}(r)$ at $J_{\rm H}=0$ (left panel) and $J_{\rm H} = 1$ (right panel), where $J_{\perp}^{zz}=0.5$ and $L=48$.
  }
\end{figure}

\begin{figure}[htbp]
  \includegraphics[width=0.65\linewidth]{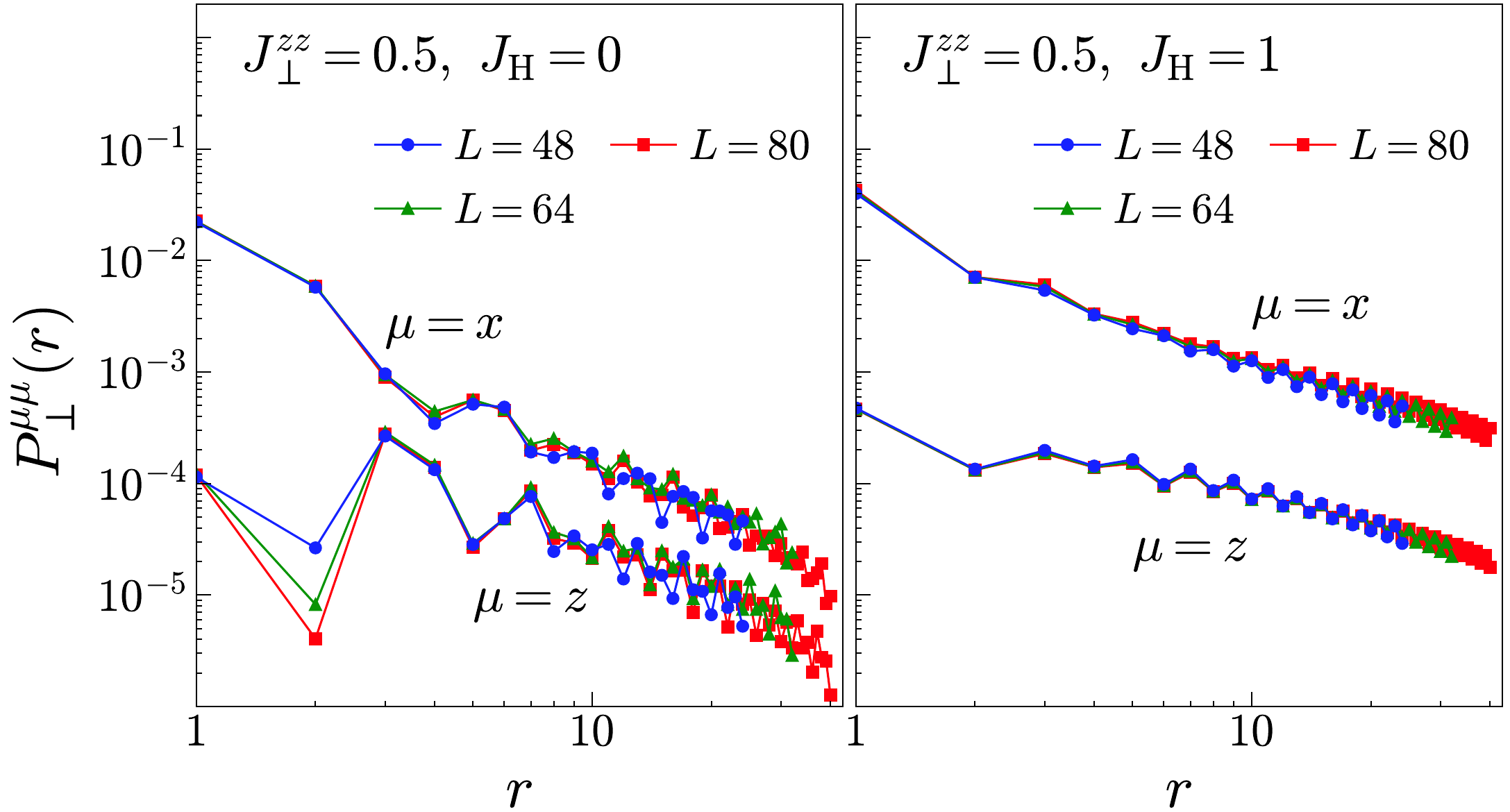}
  \caption{\label{fig:pair-corr_L-dependence}
  $L$ dependence of $P_{\perp}^{xx}(r)$ and $P_{\perp}^{zz}(r)$ at $J_{\rm H}=0$ (left panel) and $J_{\rm H} = 1$ (right panel), where $J_{\perp}^{zz}=0.5$. Note that we choose the reference site as $j_{\rm ref}=L/4$ for each $L$.
  }
\end{figure}

To confirm the validity of our result at the bond dimension $m = 10000$ and the length $L=48$ used in the main text, we plot the pair correlation function $P^{\mu\nu}_{\perp}(r)$ with varying the values of $m$ and $L$ in Figs.~\ref{fig:pair-corr_m-dependence} and \ref{fig:pair-corr_L-dependence}, where we compare the results at two sets of parameters for $\hat{H}_J$: $J_{\parallel}^{xx}=0.5$, $J_{\perp}^{zz}=0.5$, and $J_{\rm H}=0$ in the left panel, and $J_{\parallel}^{xx}=0.5$, $J_{\perp}^{zz}=0.5$, and $J_{\rm H}=1$ in the right panel.
In Fig.~\ref{fig:pair-corr_m-dependence}, we show the $m$ dependence of the pair correlation functions at $L=48$. 
In these parameter sets, no significant changes are observed in the local electron densities and pair correlations at $L=48$ when $m \geq 6000$.
In Fig.~\ref{fig:pair-corr_L-dependence}, we show the $L$ dependence of $P^{\mu\nu}_{\perp}(r)$ at $m=10000$. 
For larger $L$, the pair correlations also exhibit a power-law decay, and the $J_{\rm H}$ dependence shows a similar tendency to that of $L=48$.


\newpage
\subsection*{B. Variability in the estimation of decay exponents $K_{\mu\nu}$}

\begin{figure}[htbp]
  \includegraphics[width=0.62\linewidth]{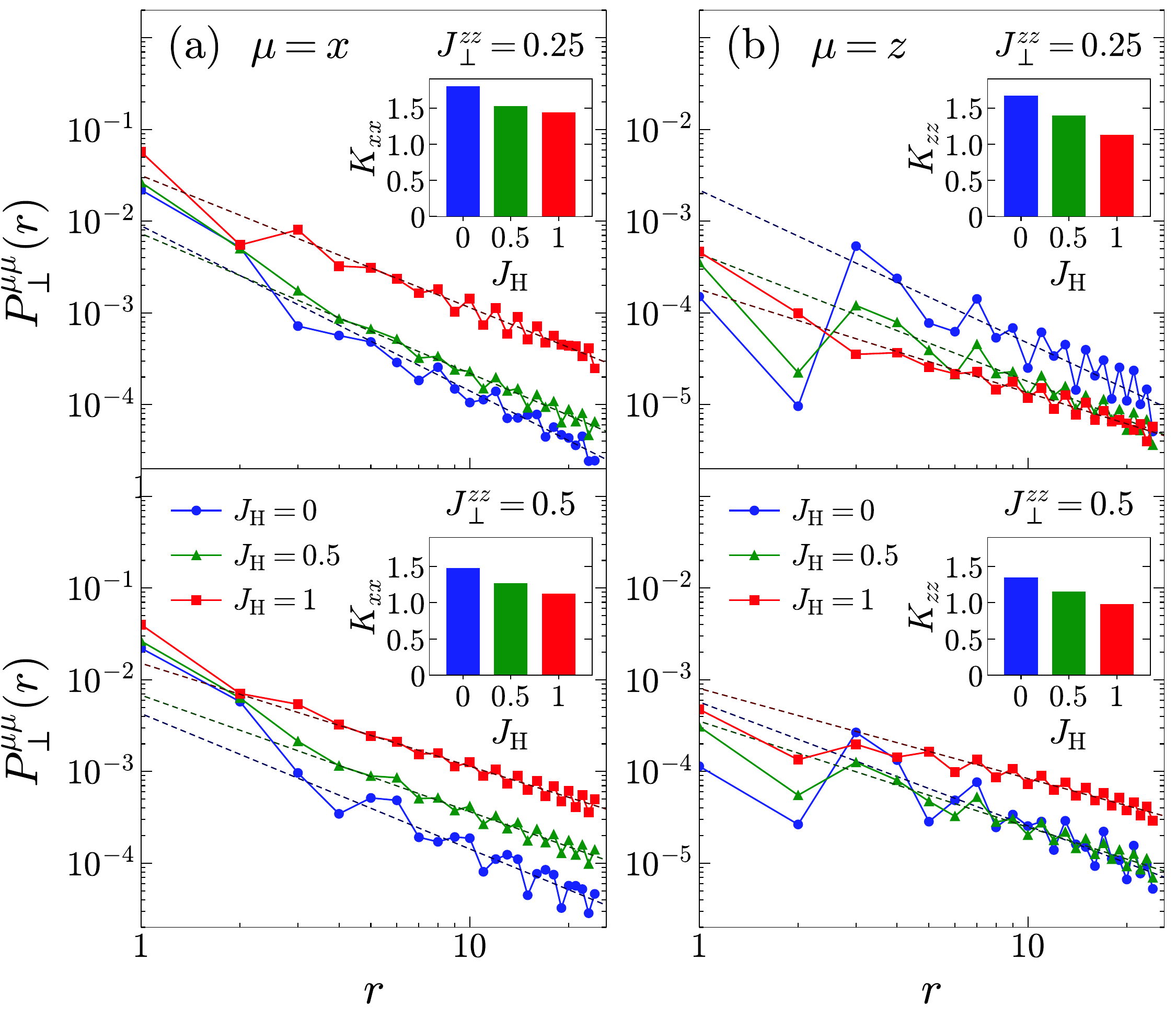}
  \caption{\label{fig:pair-corr_Hund-dependence_different_fit}
  Pair correlation functions $P_{\perp}^{\mu\mu}(r)$ for various values of Hund's coupling $J_{\rm H}$.
  (a)~$P_{\perp}^{xx}(r)$ at $J_{\perp}^{zz} = 0.25$ (upper panel) and $J_{\perp}^{zz} = 0.5$ (lower panel).  
  (b)~$P_{\perp}^{zz}(r)$ at $J_{\perp}^{zz} = 0.25$ (upper panel) and $J_{\perp}^{zz} = 0.5$ (lower panel). 
  The insets show the decay exponents of $P_{\perp}^{\mu\mu}(r)$, where the exponent $K_{\mu\mu}$ is extracted by fitting all data points at $r \ge 8$.}
\end{figure}

\begin{figure}[htbp]
  \includegraphics[width=0.62\linewidth]{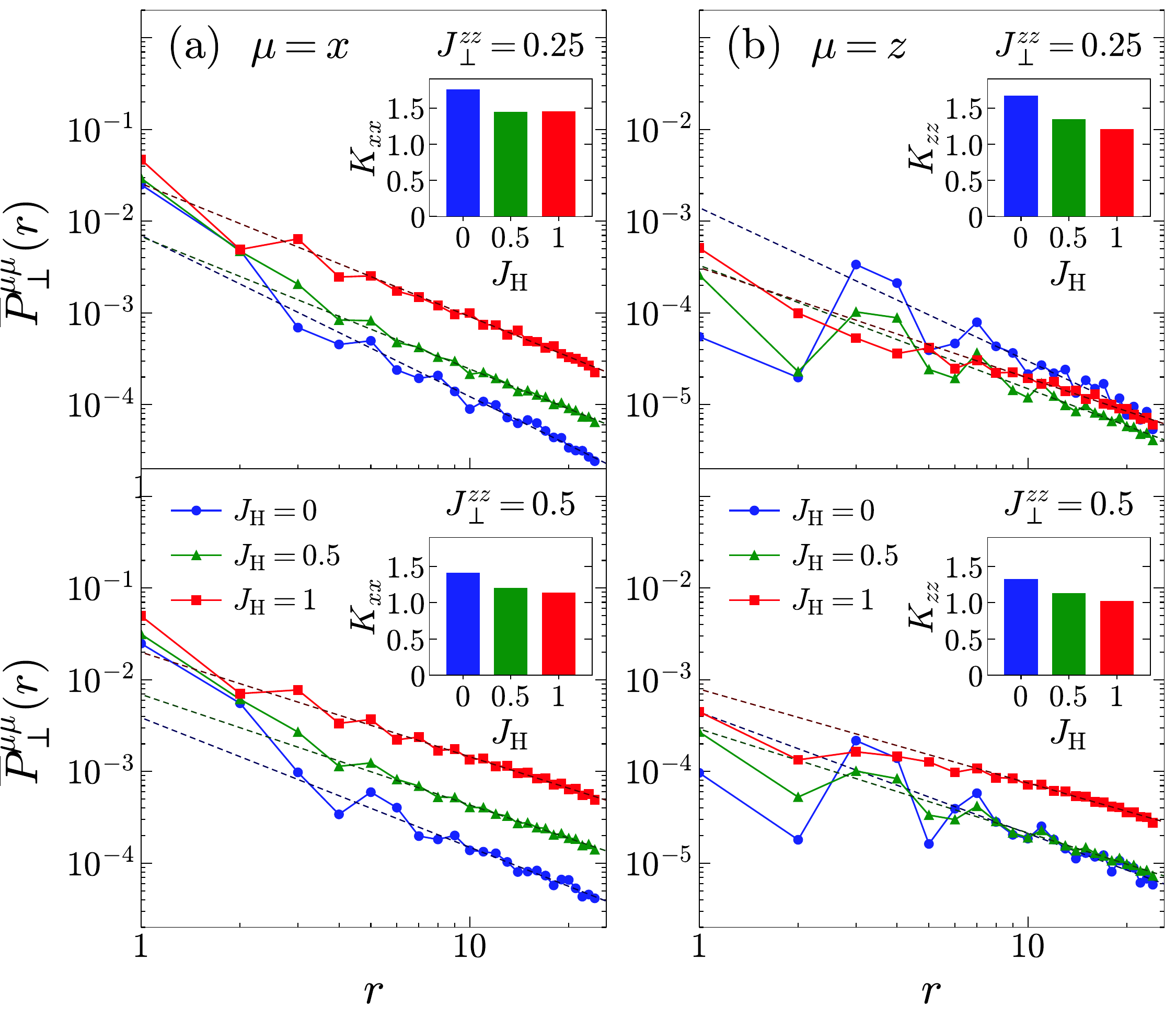}
  \caption{\label{fig:pair-corr-avg_Hund-dependence}
  Averaged pair correlation functions $\bar{P}_{\perp}^{\mu\mu}(r)$ for various values of Hund's coupling $J_{\rm H}$.
  (a) $\bar{P}_{\perp}^{xx}(r)$ at $J_{\perp}^{zz} = 0.25$ (upper panel) and $J_{\perp}^{zz} = 0.5$ (lower panel). 
  (b) $\bar{P}_{\perp}^{zz}(r)$ at  $J_{\perp}^{zz} = 0.25$ (upper panel) and $J_{\perp}^{zz} = 0.5$ (lower panel).  
  The insets show the decay exponents of $\bar{P}_{\perp}^{\mu\mu}(r)$, where the exponent $K_{\mu\mu}$ is extracted by fitting all data points at $r \ge 8$.
  }
\end{figure}

\begin{figure}[htbp]
  \includegraphics[width=0.62\linewidth]{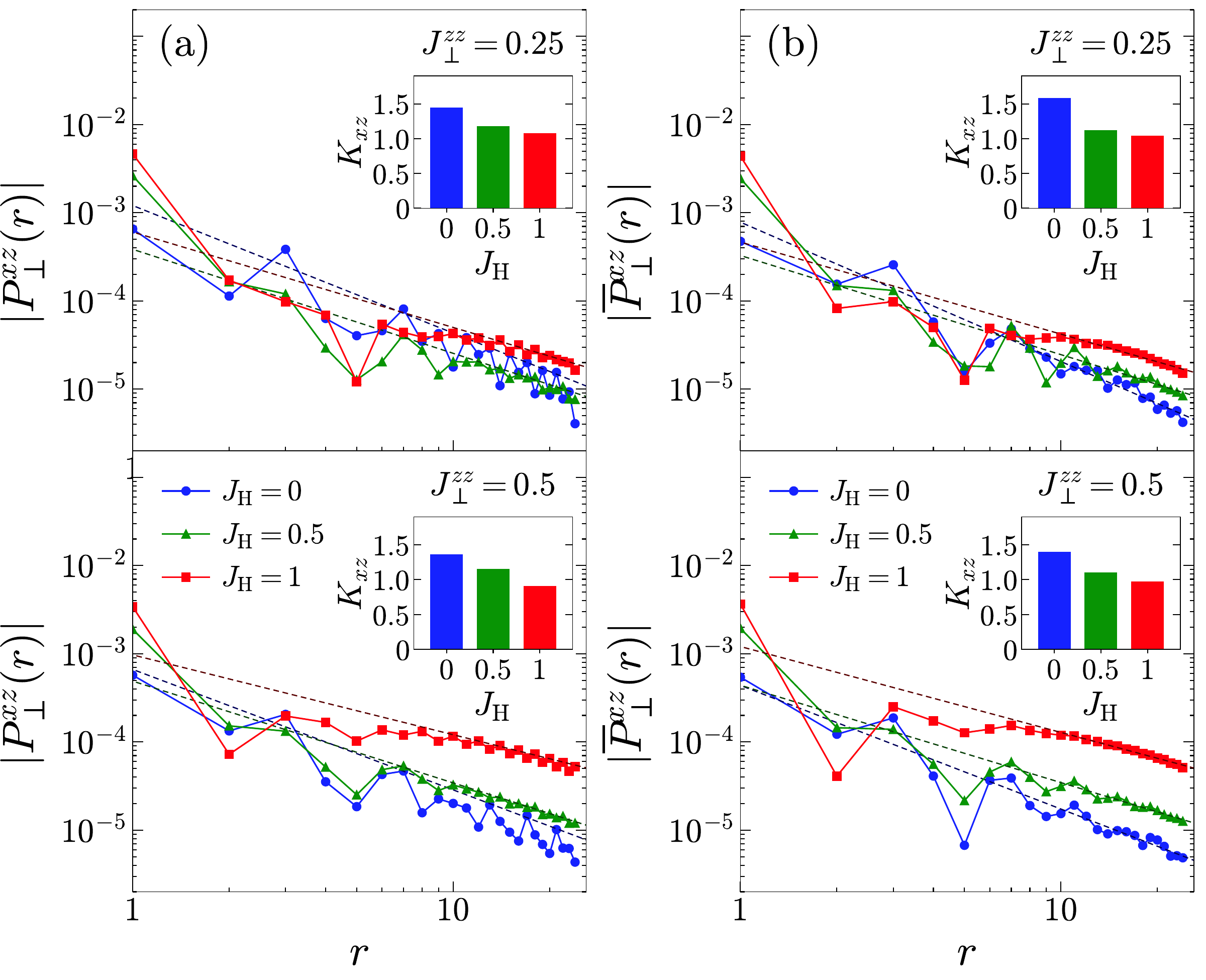}
  \caption{\label{fig:pair-corr_Hund-dependence_xz}
  (a) Interorbital pair correlation functions $P_{\perp}^{xz}(r)$ for various values of Hund's coupling $J_{\rm H}$ at $J_{\perp}^{zz} = 0.25$ (upper panel) and $J_{\perp}^{zz} = 0.5$ (lower panel). 
  The insets show the decay exponents of $P_{\perp}^{xz}(r)$, where the exponent $K_{xz}$ is extracted by fitting the crests of the data points at $r \ge 6$.
  (b) Averaged interorbital pair correlation functions $\bar{P}_{\perp}^{xz}(r)$ for various values of  $J_{\rm H}$ at $J_{\perp}^{zz} = 0.25$ (upper panel) and $J_{\perp}^{zz} = 0.5$ (lower panel). 
  The insets show the decay exponents of $\bar{P}_{\perp}^{xz}(r)$, where the exponent $K_{xz}$ is extracted by fitting all data points at $r \ge 10$.
  Note that $P_{\perp}^{xz}(r)$ and $\bar{P}_{\perp}^{xz}(r)$ have negative values at $r=2$ for each parameters.
  }
\end{figure}

The quantitative estimation of the decay exponents $K_{\mu\nu}$ can be influenced by fitting procedures such as the choice of data points for fitting.
In the insets of Fig.~\ref{fig:pair-corr_Hund-dependence_different_fit}, we show the decay exponents $K_{\mu\mu}$ extracted by fitting all data points at $r \ge 8$.
While there is little quantitative variance in $K_{\mu\mu}$, a similar tendency to those in Fig.~3 in the main text is obtained across different fitting procedures.
Also, the values of $P_{\perp}^{\mu\nu}(r)$ depend on the choice of $j_{\rm ref}$~\cite{Dolfi_2015}. 
Therefore, we also examine the dependence on $J_{\perp}^{zz}$ and $J_{\rm H}$ using the averaged pair correlation function defined by
\begin{equation}
    \bar{P}_{\perp}^{\mu\nu}(r) = \frac16 \sum_{s=0}^{5} \left\langle\Delta_{j_0+s,\mu\nu}^{\dag}\Delta^{}_{j_0+s+r,\mu\nu}\right\rangle
\end{equation}
with $j_0 = (L-r+1)/2$ if $r$ is odd and $j_0 = (L-r+2)/2$ if $r$ is even.
In Fig.~\ref{fig:pair-corr-avg_Hund-dependence}, we plot $\bar{P}^{\mu\mu}_{\perp}(r)$ for the $\mu=x$ and $z$ orbitals.
The averaged $\bar{P}^{\mu\mu}_{\perp}(r)$ and its decay exponent $K_{\mu\mu}$ show the consistent tendencies with the single-$j_{\rm ref}$ pair correlation function $P^{\mu\mu}_{\perp} (r)$ shown in Fig.~3 in the main text, where $J^{zz}_{\perp}$ and $J_{\rm H}$ promote a slow decay of the pair correlation. 
In Fig.~\ref{fig:pair-corr_Hund-dependence_xz}, we compare the single-$j_{\rm ref}$ correlation function $P^{xz}_{\perp}(r)$ and averaged correlation function $\bar{P}^{xz}_{\perp}(r)$ for the interorbital ($x$-$z$) pairs. 
Both $P^{xz}_{\perp}(r)$ and $\bar{P}^{xz}_{\perp}(r)$ show qualitatively consistent dependencies on $J_{\perp}^{zz}$ and $J_{\rm H}$. 
In both correlation functions, the decay exponent $K_{xz}$ reaches the smallest value at $J_{\perp}^{zz}=0.5$ and $J_{\rm H} = 1$. 

\clearpage

\subsection*{C. Effects of $t_{\perp}^{zz}$ and $\Delta E$}

\begin{figure}[htbp]
  \includegraphics[width=0.9\linewidth]{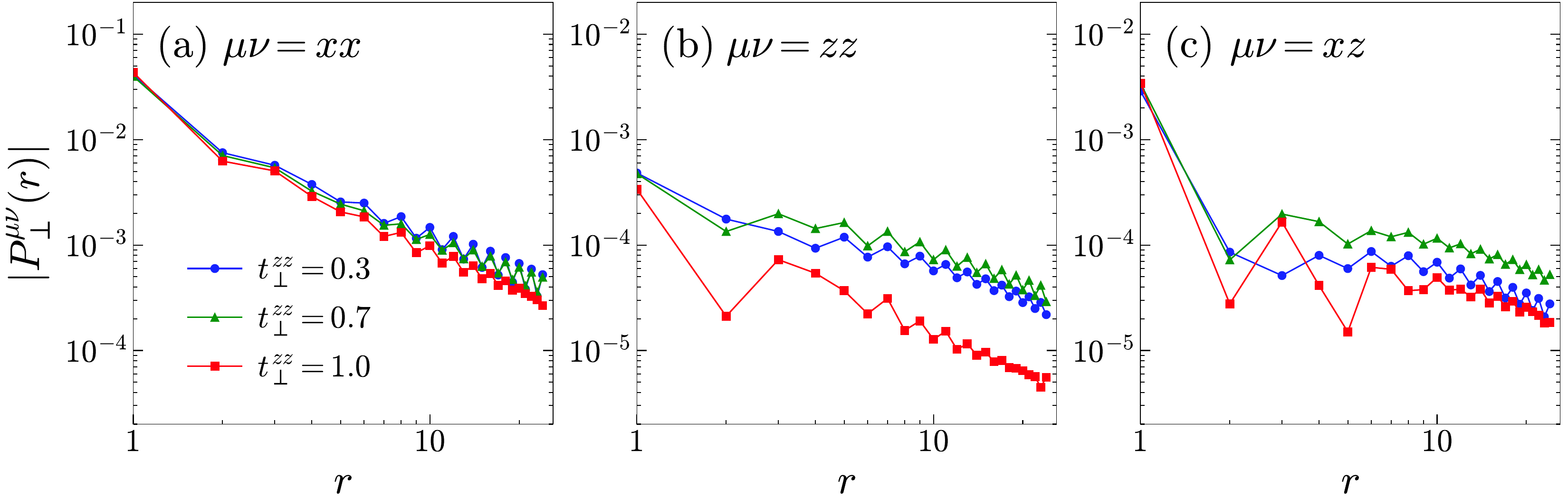}
  \caption{
    Pair correlation functions (a) $P_{\perp}^{xx}(r)$, (b) $P_{\perp}^{zz}(r)$, and (c) $P_{\perp}^{xz}(r)$ for various values of the interchain hopping $t_{\perp}^{zz}$ (where $J_{\perp}^{zz}=0.5$ and $J_{\rm H} = 1$). 
  Note that $P_{\perp}^{xz}(r)$ has a negative value at $r=2$ for each $t_{\perp}^{zz}$.
  }
 \label{fig:pair-corr_t_perp-dependence_xx-zz-xz}  
\end{figure}

\begin{figure}[htbp]
  \includegraphics[width=0.85\linewidth]{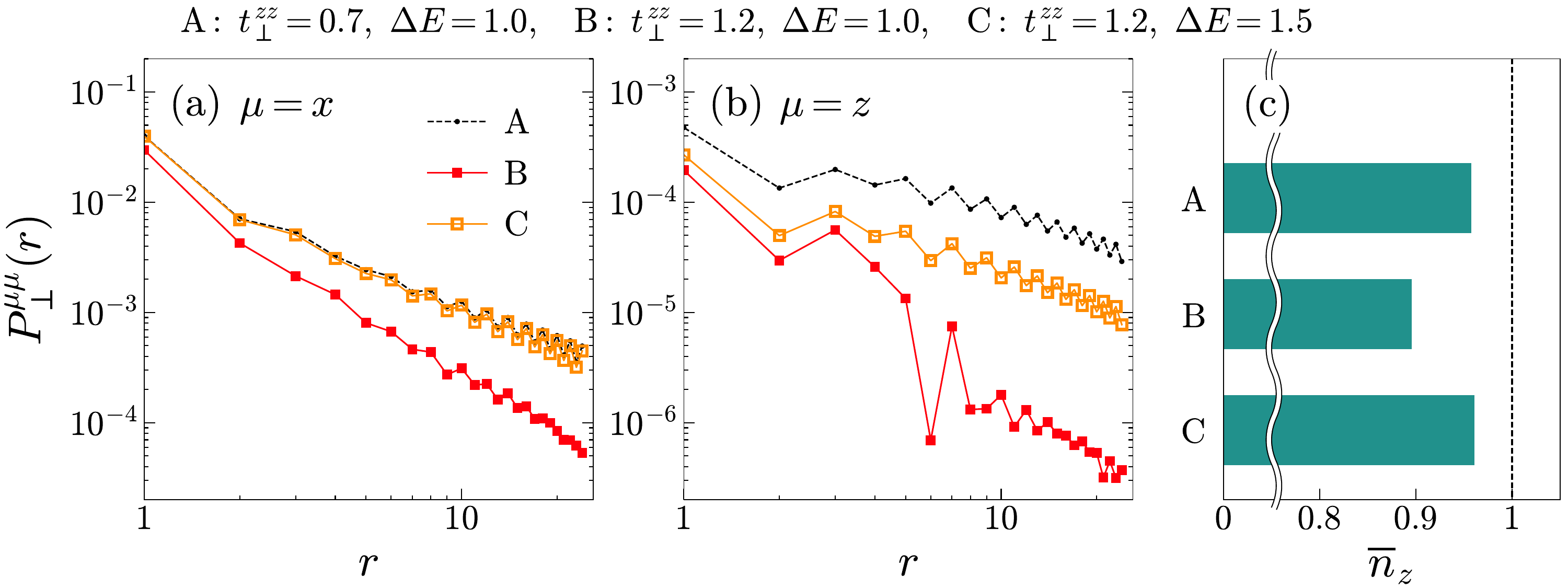}
  \caption{
    Pair correlation functions (a) $P_{\perp}^{xx}(r)$ and (b) $P_{\perp}^{zz}$ for various values of the interchain hopping $t_{\perp}^{zz}$ and the energy level difference $\Delta E$ (where $J_{\perp}^{zz}=0.5$ and $J_{\rm H} = 1$).
    (c) Electron filling of the $z$ orbital $\bar{n}_z=1/(2L)\sum_{j,l} \langle \hat{n}_{j,l,z} \rangle$. 
    The $z$ orbital is exactly half-filled when $\bar{n}_z=1$. 
  }\label{fig:pair-corr_dE-dependence}
\end{figure}

In contrast to the magnetic coupling $J_{\perp}^{zz}$, the interchain hopping $t_{\perp}^{zz}$ in the $t$-$J$ model potentially suppresses the superconducting tendency because the move of carriers across rungs possibly disturbs the development of the pair correlations along the chain direction~\cite{Hirthe_2023}.
In Fig.~\ref{fig:pair-corr_t_perp-dependence_xx-zz-xz}, we plot the pair correlation function for various values of $t_{\perp}^{zz}$. 
Although the magnitude of $P_{\perp}^{zz}(r)$ is reduced, a power-law decay is maintained and the decay exponent $K_{zz}$ is still less than 2 at $t_{\perp}^{zz}=1$. 
On the other hand, since the parameter of the $x$ orbital is not modified, the change of $t_{\perp}^{zz}$ does not suppress $P_{\perp}^{xx}(r)$ drastically. 
The magnitude of $P_{\perp}^{zx}(r)$ is slightly suppressed by $t_{\perp}^{zz}$, but the decay tendency is not so modified. 
When the interchain hopping is up to $t_{\perp}^{zz}=1.2$, the pair correlations are strongly reduced by the effect of $t_{\perp}^{zz}$.  
However, tuning the energy level difference $\Delta E$ to adjust the electron filling comparable to that at $t_{\perp}^{zz}=0.7$ [see Fig.~\ref{fig:pair-corr_dE-dependence}(c)], the decaying behaviors of the pair correlations are recovered. 

\newpage
\subsection*{D. Other correlation functions}

\begin{figure}[htbp]
  \includegraphics[width=0.9\linewidth]{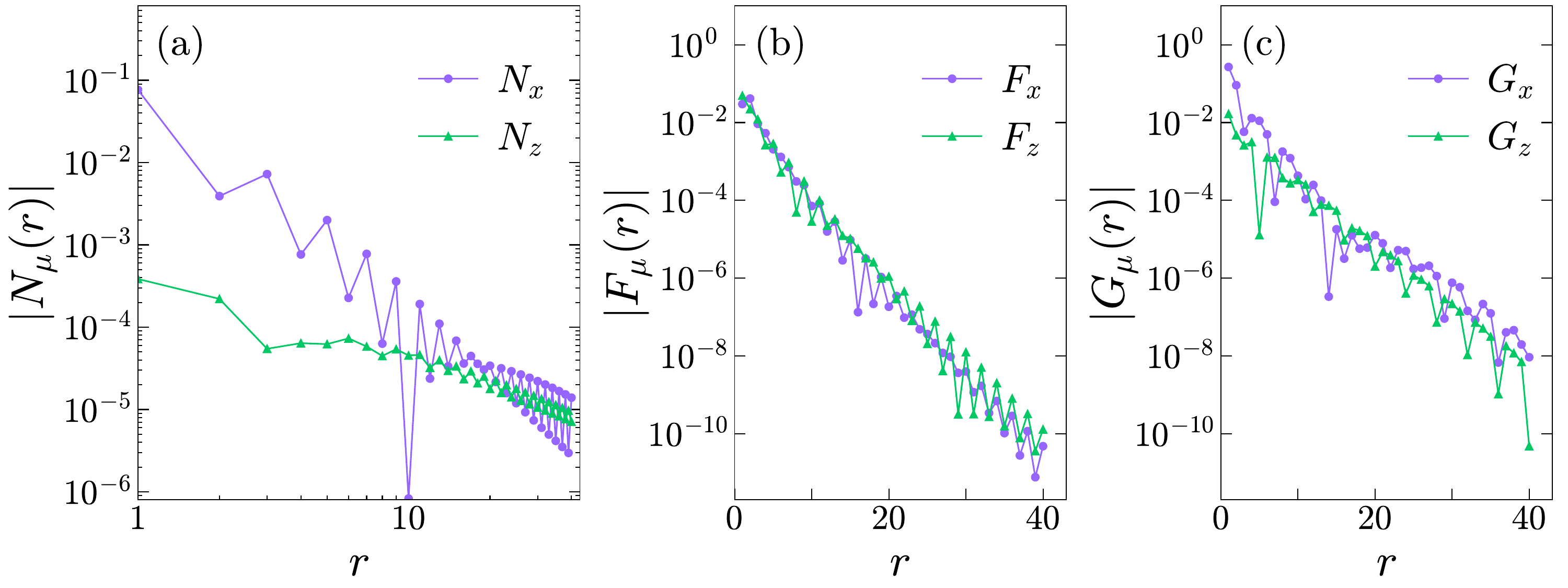}
  \caption{\label{fig:correlations}
  (a)~Charge correlation functions $N_{\mu}(r)$, (b)~spin correlation functions $F_{\mu}(r)$, and (c)~single-particle Green's functions $G_{\mu}(r)$ at $J_{\perp}^{zz}=0.5$ and $J_{\rm H}=1$ with the system length $L=80$.}
\end{figure}

We have shown that a quasi-long-range interchain pairing correlation develops in the two-orbital $t$-$J$ ladder model. 
However, the system is expected to actually become superconducting when the pair correlation dominates over other correlations.
To understand the details of ground-state properties in the two-orbital $t$-$J$ ladder, we examine various intrachain correlation functions: charge correlation function
\begin{equation}
    N_{\mu}(r) = \frac12 \sum_{l} \langle \hat{n}_{j,l,\mu}\hat{n}_{j+r,l,\mu} \rangle - \langle \hat{n}_{j,l,\mu} \rangle \langle \hat{n}_{j+r,l,\mu} \rangle,
\end{equation}
spin correlation function
\begin{equation}
    F_{\mu}(r) = \frac12 \sum_{l} \left\langle \hat{\bm{S}}_{j,l,\mu} \cdot \hat{\bm{S}}_{j+r,l,\mu} \right\rangle,
\end{equation}
and single-particle Green's function
\begin{equation}
    G_{\mu}(r) = \frac12 \sum_{l,\sigma} \left\langle \hat{\tilde{c}}^{\dag}_{j,l,\mu,\sigma} \hat{\tilde{c}}^{}_{j+r,l,\mu,\sigma} \right\rangle. 
\end{equation}
In Fig.~\ref{fig:correlations}, we show these correlations at $J_{\parallel}^{xx}=0.5$, $J_{\perp}^{zz}=0.5$, and $J_{\rm H}=1$, where $L=80$ and we set $j=j_{\rm ref}=L/4$ as the reference site. 
As shown in Figs.~\ref{fig:correlations}(b) and \ref{fig:correlations}(c), both spin correlations and single-particle Green's functions exhibit exponential decays for both $x$ and $z$ orbitals, which suggest the presence of gaps in spin and single-particle excitations~\cite{X-Lu_2023}.
Similar decaying behaviors of the correlation functions are also observed in the single-orbital $t$-$J$ ladder model~\cite{X-Lu_2023}.
On the other hand, in Fig.~\ref{fig:correlations}(a), we find that the charge correlations exhibit a power-law decay.
The decay rates of the charge correlations for both orbitals are similar at long distances. 
The local electron density $n_{\mu}(j)$ does not show a substantial charge-density-wave (CDW) like character around the center of the ladder as shown in Fig.~2 in the main text, and the decays of the charge correlations are faster than those of the interchain pair correlations $P^{\mu\mu}_{\perp}(r)$.
Hence, we may not expect a CDW in the parameter set used in Fig.~\ref{fig:correlations}.  

\begin{figure}[htbp]
  \includegraphics[width=0.4\linewidth]{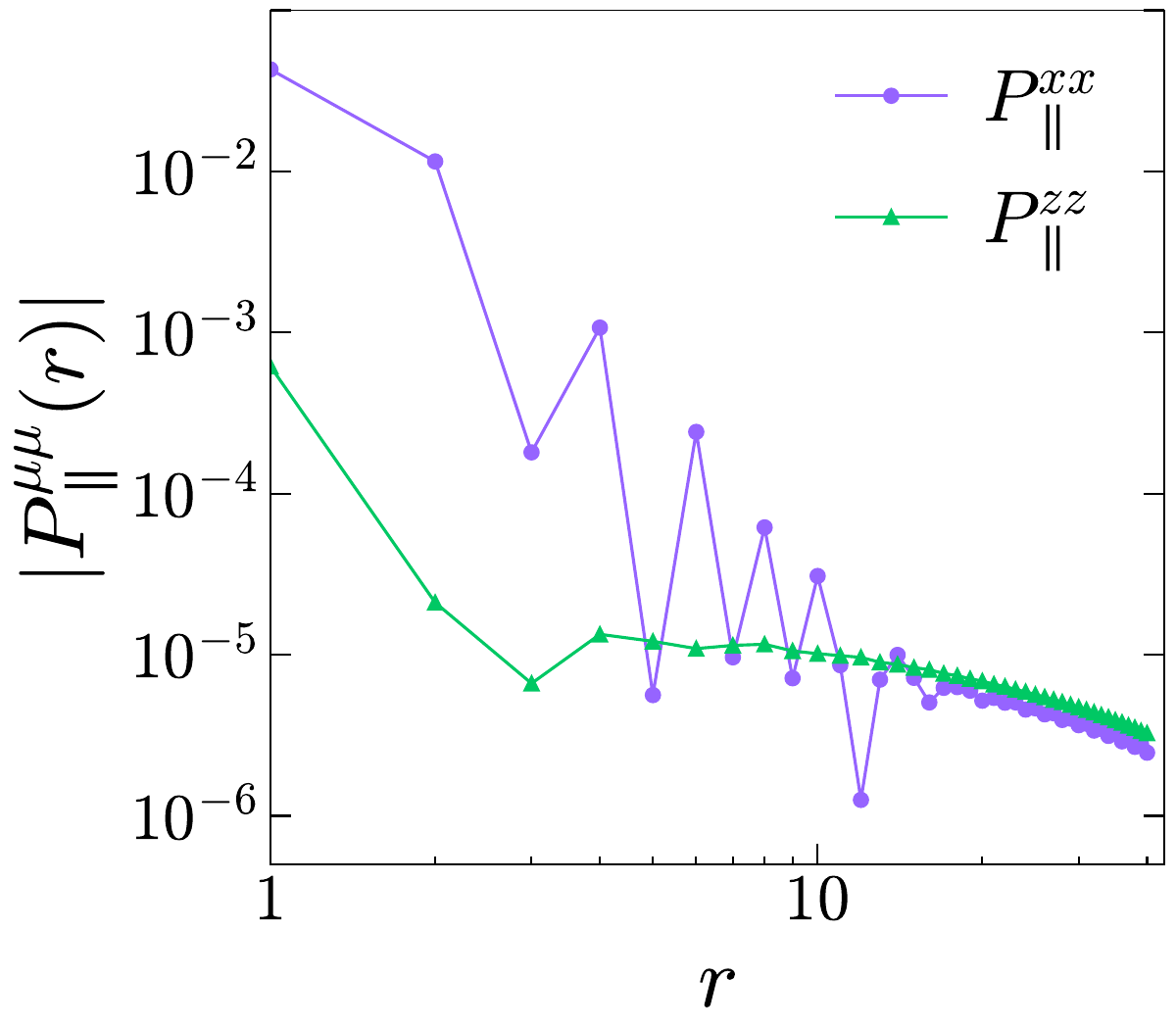}
  \caption{\label{fig:pair-corr_intrachain}
  Intrachain pair correlation functions $P_{\parallel}^{\mu\mu}(r)$ at $J_{\perp}^{zz}=0.5$ and $J_{\rm H}=1$ with the system length $L=80$.}
\end{figure}

We also examine the intrachain pair correlation
\begin{equation}
    P_{\parallel}^{\mu\nu}(r) = \frac12 \sum_{l} \left\langle \hat{\Delta}^{\dag}_{\parallel\,j,l,\mu\nu} \hat{\Delta}^{}_{\parallel\,j+r,l,\mu\nu} \right\rangle,
\end{equation}
where
\begin{equation}
    \hat{\Delta}_{\parallel\,j,l,\mu\nu} = \frac{1}{\sqrt{2}} \left(\hat{\tilde{c}}^{}_{j,l,\mu,\uparrow}\hat{\tilde{c}}^{}_{j+1,l,\nu,\downarrow}-\hat{\tilde{c}}^{}_{j,l,\mu,\downarrow}\hat{\tilde{c}}^{}_{j+1,l,\nu,\uparrow}\right)
\end{equation}
is the intrachain spin-singlet pair annihilation operator.
As shown in Fig.~\ref{fig:pair-corr_intrachain}, the intrachain pair correlations for both orbitals exhibit power-law decays whose decay rates at long distances are comparable to interchain ones. This may result from the hybridized entity obtaining a quasi-long-range correlation.
Nevertheless, the absolute values of $P^{\mu\nu}_{\parallel}(r)$ are smaller than those of $P^{\mu\nu}_{\perp}(r)$, which implies the intrachain pairs are minor contributors to the pairing in the hybridized entity.


\end{document}